\documentclass[10pt,letterpaper]{article}

\usepackage{cogsci}
\usepackage{graphicx}
\usepackage{amsmath}

\cogscifinalcopy 

\usepackage{pslatex}
\usepackage{apacite}
\usepackage{float} 

\title{Visual and Auditory Aesthetic Preferences Across Cultures}
 
\author{
    {\large \bf Harin Lee\textsuperscript{1,2,*}, Eline Van Geert\textsuperscript{3}, Elif Çelen\textsuperscript{1}, Raja Marjieh\textsuperscript{4}, Pol van Rijn\textsuperscript{1}, Minsu Park\textsuperscript{5}, and Nori Jacoby\textsuperscript{1,6}} \\
  \textsuperscript{1}Max Planck Institute for Empirical Aesthetics, Frankfurt am Main, Germany \\
  \textsuperscript{2}Max Planck Institute for Human Cognitive and Brain Sciences, Leipzig, Germany \\
  \textsuperscript{3}KU Leuven, Leuven, Belgium \\
  \textsuperscript{4}Princeton University, New Jersey, United States \\
  \textsuperscript{5}New York University Abu Dhabi, Abu Dhabi, United Arab Emirates \\
  \textsuperscript{6}Cornell University, New York, United States \\ \\
  *Corresponding author email: \url{harin.lee@ae.mpg.de} \\
}

\begin{document}
\maketitle

\begin{abstract}
Research on how humans perceive aesthetics in shapes, colours, and music has predominantly focused on Western populations, limiting our understanding of how cultural environments shape aesthetic preferences. We present a large-scale cross-cultural study examining aesthetic preferences across five distinct modalities extensively explored in the literature: shape, curvature, colour, musical harmony and melody. We gather 401,403 preference judgements from 4,835 participants across 10 countries, systematically sampling two-dimensional parameter spaces for each modality. The findings reveal both universal patterns and cultural variations. Preferences for shape and curvature cross-culturally demonstrate a consistent preference for symmetrical forms. While colour preferences are categorically consistent, ratio-like preferences vary across cultures. Musical harmony shows strong agreement in interval relationships despite differing regions of preference within the broad frequency spectrum, while melody shows the highest cross-cultural variation. These results suggest that aesthetic preferences emerge from an interplay between shared perceptual mechanisms and cultural learning.


\end{abstract}

\section{Introduction}
Our everyday lives are filled with aesthetic experiences. When choosing the outfit to wear, we consider colour combinations. When listening to music, certain sounds evoke pleasure while others create discomfort. These aesthetic judgements, deeply personal, are embedded within broader cultural frameworks that shape our taste and experiences of the world~\cite{che2018cross}.

The origin of aesthetic preferences has been debated throughout history, from Plato's philosophical discourse on beauty~\cite{pappas_platos_2024} to contemporary cognitive scientists~\cite{palmer_visual_2013}. However, it remains open to what extent aesthetic preferences arise from cultural learning (preferences arising from exposure to specific cultural styles, conventions, and norms), or from biologically rooted universal principles (preferences arising from fundamental properties of the nervous system, sensory processing, or basic cognitive mechanisms that are largely consistent across individuals and cultures~\citeNP{che2018cross,sharman1997anthropology,bertamini2020}).

Previous research, predominantly focused on Western populations, has offered limited insight into this question about their cross-cultural generalisability~\cite{blasi_over-reliance_2022, che2018cross}. We address this gap through a large-scale (N = 4,835), cross-cultural (10 countries) investigation of aesthetic preferences across five distinct modalities: shape (aspect ratio of rectangle), curvature (Bézier curve; \citeNP{Farin1993-ac}), colour (differing in hue degrees), musical harmony and melody (pitch intervals). Notably, we focus on two-dimensional representations of these five distinct modalities so that they can be continuously sampled without relying on discrete categories (e.g., canonical aspect ratios in shape and curvature). This is in contrast with many existing studies that rely on a predefined set of stimuli, enabling to capture controlled parametric variations.

Here, we aim to identify both universal patterns, potentially reflecting innate preferences stemming from fundamental aspects of human perception (e.g., symmetry processing or basic sensory consonance), and culturally specific variations, shaped by exposure to diverse artistic traditions and cultural norms. We hypothesise that certain modalities, such as shape and curvature, may exhibit greater universality due to shared perceptual biases~\cite{bertamini2020, little_preferences_2007}, while others, such as melody and color combinations, may demonstrate more cultural variability reflecting the influence of learnt associations and cultural conventions~\cite{mcdermott_indifference_2016, anglada-tort_large-scale_2023}.

\begin{figure*}[t] 
  \centering
    \includegraphics[width=\textwidth]{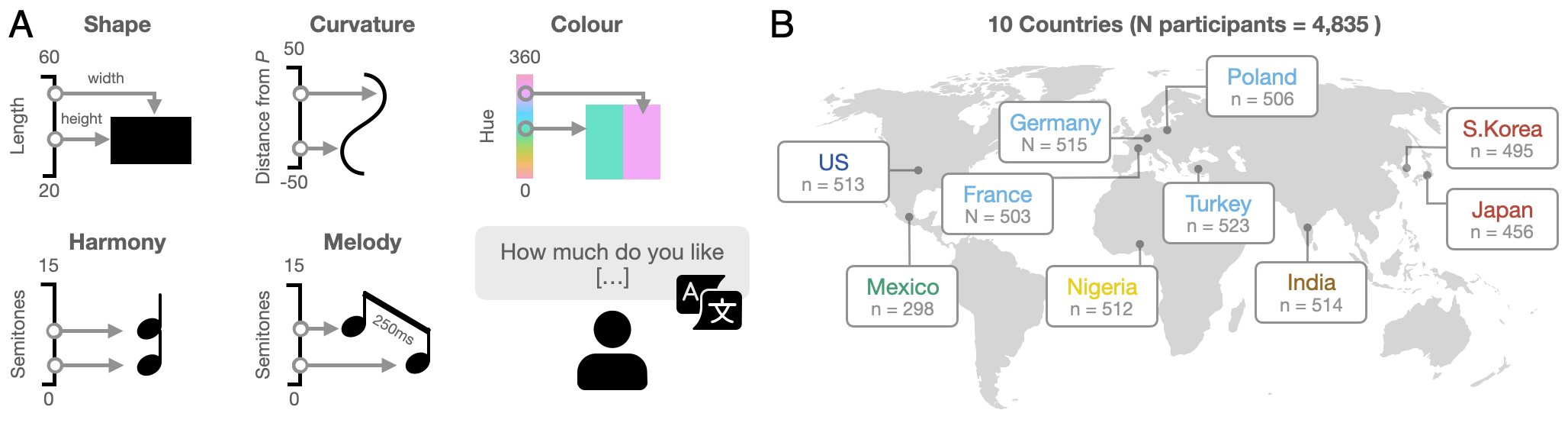} 
    \caption{
      Schematics of experimental design. 
      (A) In independent experiments for each modality, participants were asked to rate how much they like the seen or heard stimulus on a 7-point scale from 1 being ``not at all'' to 7 ``very much'', translated into their own language. Each presented stimulus was defined at random by sampling two points from the general space (see `Defining Stimulus Space' in Methods).
      (B) Participants were recruited from 10 countries, including all continents which are coloured according to regions defined by WorldBank (\url{www.worldbank.org}).
    } 
  \label{fig:fig1} 
  \end{figure*}
    
\section{Background}
We deliberately selected modalities that have been extensively explored in empirical aesthetics research for comparisons with our cross-cultural insights. Below, we summarise key findings on these modalities.

\subsection{Shape}
Early studies have shown that rectangular shapes following specific ratio rule, such as the \textit{golden ratio} (approximately 1:1.618), are often perceived as more aesthetically appealing~\cite{fechner1876}.
However, individual differences in aesthetic preferences for rectangles have emerged across studies~\cite{green_all_1995, mcmanus2010, mcmanus_square_2013}, and these variations challenge the assumed universality of the golden ratio~\cite{stieger_time_2015}. Other studies have indicated the brain processes horizontal proportions with greater ease (thus more aesthetically appealing), as this scanning direction aligns naturally with our landscape-oriented vision system~\cite{mcmanus_square_2013}.

\subsection{Curvature}
Curved shapes have consistently been shown to elicit more favourable responses, being perceived as pleasant, calming and beautiful, compared to angular forms~\cite{chuquichambi_how_2022}. Smooth curvature, characterised by gradual transitions along lines or surfaces, has shown to induce greater aesthetic appeal than abrupt or angular changes. This concept is notably illustrated by William Hogarth's \textit{line of beauty}~\cite{hogarth_analysis_1753}, an S-shaped curve that embodies ideal curvature. Hogarth posited that moderate curvature strikes the optimal aesthetic balance, avoiding both excessive flatness and extreme undulation.

\subsection{Colour}
Preferences for colour combinations are understood to emerge from a complex interplay of ecological associations, harmony principles, and cultural or personal contexts (for meta-review, see~\citeNP{palmer_visual_2013}). Contrasting colours can create compelling visual effects, particularly when warm figures (e.g., red or yellow) appear against cool backgrounds (e.g., blue or green), enhancing perceived saturation~\cite{schloss_aesthetic_2011}. Notable cross-cultural differences have been observed in colour preferences~\cite{taylor_color_2013}, which may stem from the distinct meaning associations that colours carry within different cultural contexts (e.g., certain colours being associated with higher status and power).

\subsection{Harmony}
Exposure to specific musical systems is thought to shape our mental representations of musical harmony. Recent cross-cultural studies involving small-scale societies suggest that perceptions of pleasantness and harmony are culturally influenced rather than universal~\cite{mcdermott_indifference_2016, McPherson2020-cb}. Moreover, a large-scale approach systematically exploring preferences for musical harmonies have revealed distinct patterns in interval preferences that are shaped by timbre~\cite{marjieh_timbral_2024}.

\subsection{Melody}
Compared to musical harmony, preferences for musical melody are relatively less understood. Experimental work explored the related areas of memory and expectations~\cite{narmour1992analysis,dowling1986music}, while corpus work identified patterns that are common in Western music~\cite{vos1989ascending,rodriguez2013perceptual}. Computational modelling approaches have utilised corpus data to understand probabilistic sequences of musical notes to predict surprise and pleasure~\cite{temperley2008probabilistic}. A recent study using iterated paradigms show that sung melodies, passed along participant chains, evolve into prototypical sequences, revealing shared internal representations and preferences~\cite{anglada-tort_large-scale_2023}.

\section{Methods}
\subsection{Cross-cultural Recruitment}
We recruited 4,835 participants (mean age across countries = 42.7, SD = 12.6) through the online CINT platform (\url{www.cint.com}), using an approved ethical protocol (Max Planck Ethics Council \#202142). The selection of countries ensured broad geographical and linguistic coverage across all continents. 
To be eligible for participation, individuals were required to have been born in their respective country and be current residents (Figure~\ref{fig:fig1}B). Participants received compensation at their country's standard wage rate. Sample sizes and specific locations were determined based on a pilot experiment conducted in English and the participant pool availability as reported by CINT.

\subsection{Defining Stimulus Space}\label{sec:define_space}
We systematically sampled parameters by defining a fixed two-parameter space for each modality. In each trial, two values were randomly selected from the given space to define the stimulus on the spot (Figure~\ref{fig:fig1}A). This way, we could uniformly sample across the entire space without discrete categories and thus pinpoint high and low preference density regions of the full continuous space. In particular, this is advantageous in a cross-cultural context, as it allows us to avoid making assumptions about specific discrete categories that are primarily derived from existing research focused on Western participants~\cite{blasi_over-reliance_2022}.

\subsubsection{Shape space.}
Rectangles were generated with varying width and height, ranging from 20 to 60 pixels, to be presented in a 100$\times$100 pixel window. The range selection represents a careful balance between detecting subtle differences in aspect ratios whilst maintaining a reasonable maximum ratio. We selected values that constrained the maximal aspect ratio between 1:1 and 1:3 (or 3:1).

\subsubsection{Curvature space.}
Smooth curved lines were created using the cubic Bézier curve formula~\cite{Farin1993-ac}:

$B(t) = (1-t)^3P_0 + 3(1-t)^2tP_1 + 3(1-t)t^2P_2 + t^3P_3$

\noindent where the curve always starts and ends at the vertical centerline ($x$=50), and $t$ ranges from 0 to 1. $P_0=(50,0)$ is the start point, $P_1=(50+a,25)$ is the first control point, $P_2 = (50+b,75)$ is the second control point, and $P_3 = (50,100)$ is the endpoint (i.e., 100px in length). The parameters $a$ and $b$ ranged from -50 to 50, controlling the curvature of the line.

\subsubsection{Colour space.}
Pairs of colours were generated using the OKLCH colour space~\cite{oklch}. Whilst there are other colour spaces such as RGB and HSV, we selected OKLCH as it more accurately represents human perception of colour similarities. We set the lightness to 70\% and chroma to 0.15, presenting two colours (25px width, 50px height) side by side with each varying hue angles from 0° to 360°. These parameters were carefully chosen to align with previous research on colour representation~\cite{schloss_aesthetic_2011, vangeert_jacoby_2024}.

\subsubsection{Harmony and melody spaces.}
Pairs of harmonic complex tones (dyads) were played simultaneously for harmony (1s in duration), while sequentially presented for melodies (750ms in duration) by including a 250ms gap, spanning a continuous MIDI range from 60 to 75 (C4 to C5). These tones were synthesised during the experiment using \textit{ToneJS} (\url{tonejs.github.io}), following the same parameter settings as~\citeNP{marjieh_timbral_2024} (10 harmonics, amplitude roll-off=12dB).

\subsection{Experimental Procedure}
Stimuli were presented using $PsyNet$~\cite{harrison2020}, a web-based experimental platform (\url{www.psynet.dev}). Each participant completed 80 trials. In each trial, participants viewed or heard a stimulus and rated ``How pleasant is this [modality type] from a scale of 1 (not at all) to 7 (very much)?'' Responses were collected using a 7-point Likert scale (Figure~\ref{fig:fig1}A). To prevent rapid clicking, participants could only proceed after a 1.5-second delay. Each trial lasted approximately 4 seconds, with the complete experiment taking 8 minutes. 

\subsubsection{Additional screening.}
For colour perception, we accounted for display variations by applying gamma correction to standardised value 1/2.2, proven effective in previous online studies~\cite{epicoco_can_2024}. Participants were instructed to disable night-shift mode and required to pass a colour blindness test consisting of six Ishihara plates~\cite{clark1924}. Participants also verified screen brightness by adjusting until three grey rectangles were visible against a darker grey background. 

For harmony and melody perception, participants were asked to adjust the volume to a comfortable level and were instructed to use headphones. An audio screening test required them to identify the odd sound amongst three options.

\section{Results}

\begin{figure*}[t] 
  \centering
    \includegraphics[width=\textwidth]{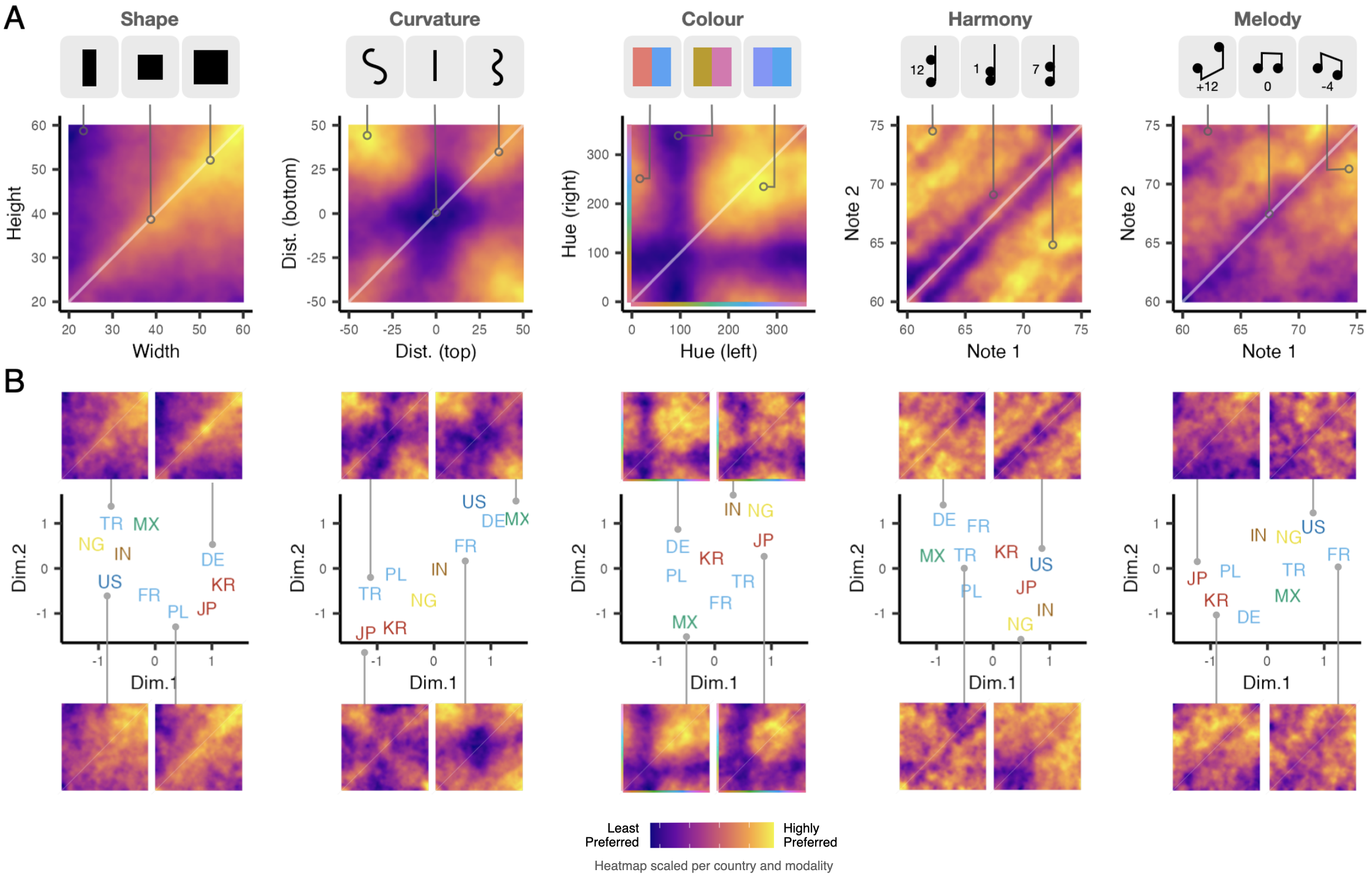} 
    \caption{
      Preferred regions in modality spaces.
      (A) Using a fixed bandwidth to smooth the preference ratings, the most preferred regions in each modality space are highlighted in yellow. White diagonal lines indicate where values above and below are equal but in differing two parameter orders.
      (B) Cross-cultural similarity in regions of preference and their variability. Jensen-Shannon distance was used to measure the similarity between country-level matrices and dimensionality reduction was performed using UMAP. Countries positioned closer together share similar regions of preference. Insets show examples from different coordinates of these UMAPs to illustrate variations.
      } 
  \label{fig:fig2} 
  \end{figure*}

\begin{figure*}[t] 
  \centering
    \includegraphics[width=\textwidth]{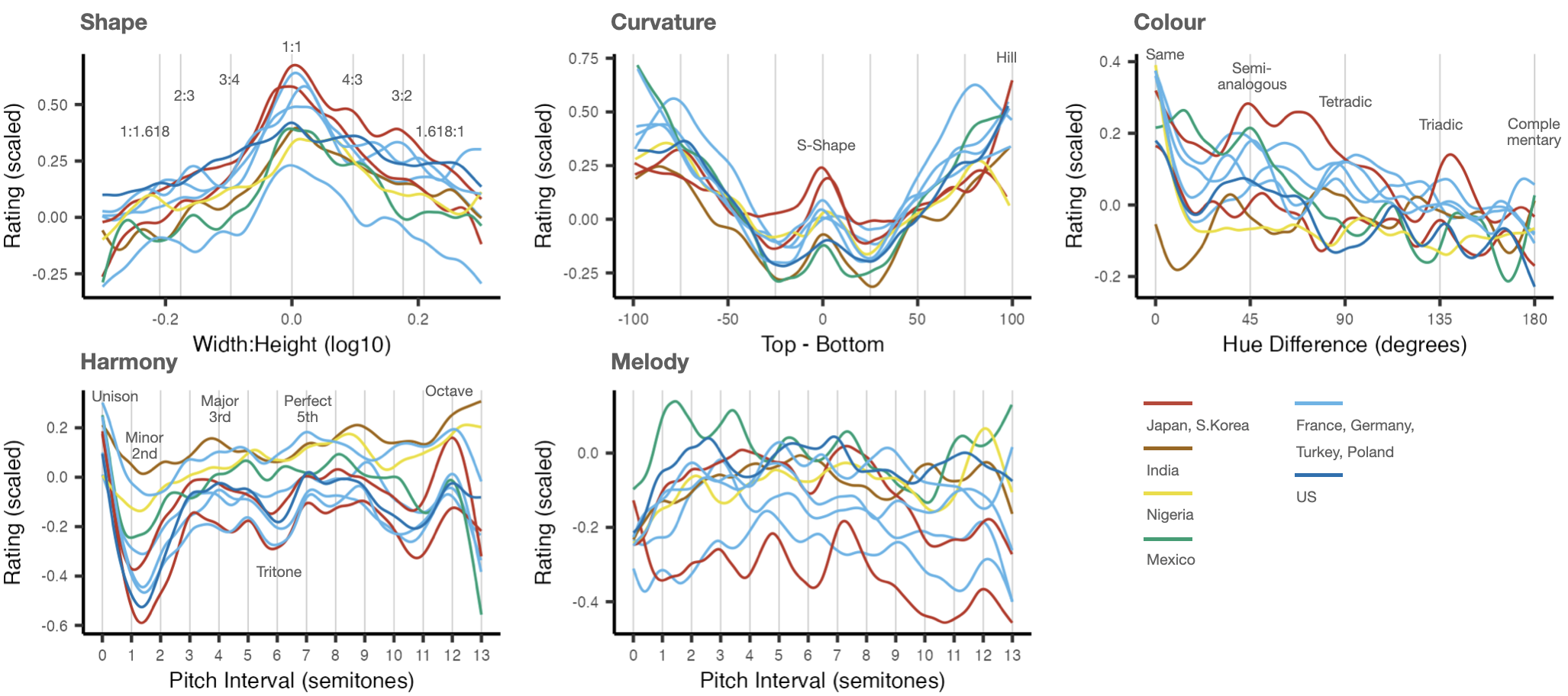} 
    \caption{
      Relational preference across modalities is assessed as follows: Shape = width-to-height aspect ratios; Curvature = the difference between control points $P1$ and $P2$; Colour = absolute difference in degrees between paired hues; Harmony and melody = pitch intervals between tone pairs in semitones. Each line represents a GAM-fitted curve per country, with colours denoting world regions.
    }
  \label{fig:fig3} 
  \end{figure*}

\begin{figure}[t] 
  \centering
    \includegraphics[width=\columnwidth]{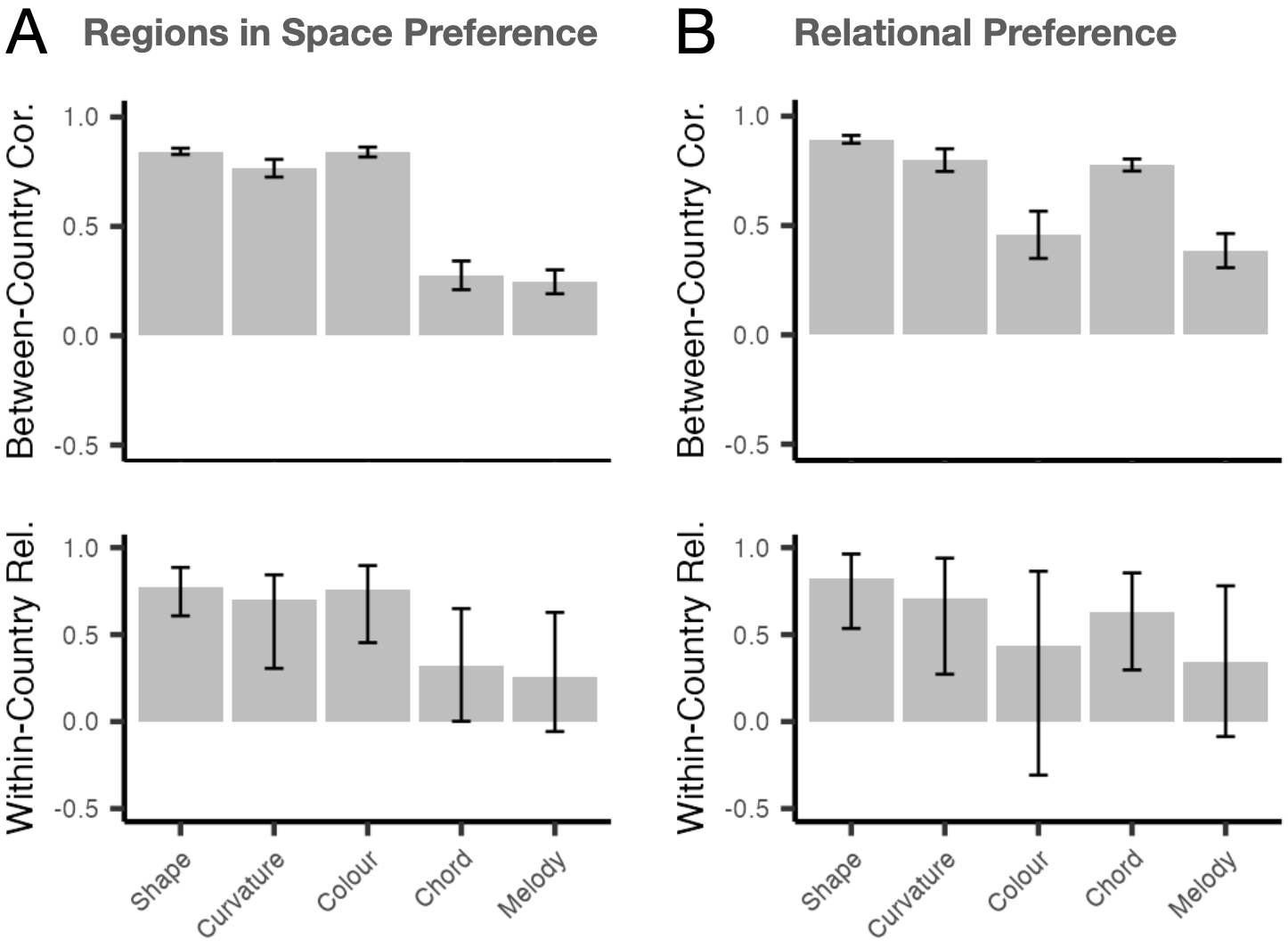} 
    \caption{
      Agreement and disagreement between cultures.
      Between-country correlations in (A) preferred regions in modality spaces, and (B) relational preferences across modalities. Below each of these, we report the reliability using split-half correlations. Error bars indicate 95\% CI.
    }
  \label{fig:fig4} 
  \end{figure}

\subsection{Preferred Regions in Modality Spaces}
We begin by exploring the regions in each modality space that participants found aesthetically pleasing (or displeasing). Figure~\ref{fig:fig2}A displays these spaces and preferred regions in yellow, aggregated across all countries. We generated these heatmaps by smoothing the preference ratings across the continuous parameter spaces with a fixed grid. Values above or below the white diagonal lines indicate the two parameters being symmetrical.

We observe that preferences in certain modalities depend strongly on the ratios between parameter values. This is evident in the strong preference following the diagonal in shape space, where perfectly squared shapes lie, and in the striped patterns in musical harmony, which reflect structured interval preferences. By contrast, other modalities show more categorical or absolute preferences. For example, colour combinations demonstrate that bluish hues are consistently preferred while dark sandy colours are disliked, regardless of their pairings. For curvature lines, the strongest preference is seen at regions that form \textit{S-shaped} patterns where $P1$ and $P2$ distances mirror each other symmetrically, but also when both control points shift equally in the same direction, creating a \textit{bumpy-hill} shape. Notably, melody preferences displays the least structured pattern, suggesting greater cross-cultural or individual variability.

To assess cross-cultural similarities in these preference spaces, we analysed each country separately and measured the similarity between them using the Jensen-Shannon distance, which is suitable for comparing probability distributions. Figure~\ref{fig:fig2}B shows these relationships after applying UMAP for dimension reduction, with various coordinates of the relation space shown as small insets.

The observed pattern reveals clear cross-cultural differences. For instance, Japanese participants uniquely prefer straight lines, a pattern not observed elsewhere---in contrast, straight lines were the least preferred in France. The striped pattern in musical harmony, seen at the global level (Figure~\ref{fig:fig2}A), is particularly prominent among German and US participants (with Germans favouring harmonies in the lower frequency spectrum), whilst it is less visible in countries such as Turkey and Nigeria.

Interestingly, geographical proximity does not necessarily indicate a similar preference. While Korea and Japan always close to each other in these preference spaces, European countries and the US show no clear grouping, suggesting that additional cultural or historical factors may be at play.

\subsection{Relational Preference}
The question of whether aesthetic appeals are governed by specific mathematical principles (i.e., ratio rules) has been extensively debated (see meta-review by~\citeNP{palmer_visual_2013}). Building on previous research exploring these structural relationships, as shown in Figure~\ref{fig:fig3}, we analysed relational (ratio-like) preferences across five modalities, including the aspect ratios of rectangular shapes, the differences between two control points of cubic Bézier curves, the differences in hue degrees between colour pairs, and pitch intervals for both harmony and melody (translated from frequency to continuous MIDI semitones). We fitted a Generalised Additive Model (GAM) with a fixed smoothing parameter ($k$ knots = 15) to capture the non-linear relationships in marginal distributions.

For rectangular shapes, we see a prominent peak at the 1:1 ratio, representing perfect squares, as was seen with concentrated preference along the diagonal in Figure~\ref{fig:fig2}. The preference strength gradually declines as the ratio deviates from this symmetrical form. Notably, we find no evidence of preference peaks at the famously known golden ratio (1.618:1 or its inverse) in any of the countries, aligning with recent observations~\cite{stieger_time_2015}. Interestingly, the data from South Korea shows distinct peaks around the 4:3 and 3:2 aspect ratios, which align with common proportions used in visual media and frame design, suggesting how these familiar aspect ratios could have appeared appealing~\cite{cossar_shape_2009}. 

A particularly intriguing finding present across all cultures is a consistent bias in participants favouring wider (z-score mean = 0.27) rectangles over taller ($m$ = 0.17) ones with equivalent aspect ratios ($t$ = 76.5, $p$ $<$ .001; seen as slight asymmetry with higher ratings above the 1:1 in Figure~\ref{fig:fig3}). This finding aligns with previous works demonstrating this bias, as objects placed wider can indicate stability and thus appear more appealing~\cite{goffaux_horizontal_2010}.

For curved lines, we observe the strongest peaks when the two control points $P1$ and $P2$ sum to zero (i.e., when they are exact opposites). Slight deviations from this symmetry result in reduced preference. Previous research mainly focused on curved lines taken from Hogarth's line of beauty~\cite{hogarth_analysis_1753, Hubner2022-pv}, which examines variations in seven categories of S-shaped curves. Our work extends beyond this finding and demonstrates that curves pointing in the same direction (e.g., difference of 100) are equally, or even more preferred than S-shaped curves. Notably, Japan and Germany uniquely demonstrate a distinct peak when the control points are roughly 75 values apart (e.g., disproportionate S-curves).

For colours, we observe a general declining trend in preference as the hue degrees between paired colours diverge, which aligns with previous observations~\cite{schloss_aesthetic_2011}. Yet, while discretisation of parameters in past works limited the resolution of this decline (though see~\citeNP{vangeert_jacoby_2024} for continuous space exploration), we see an apparent dip at approximately 30° degrees difference. This suggests visual dissonance when colours are similar but slightly mismatched. Several countries exhibit preference peaks at around 45° degrees, corresponding to \textit{semi-analogous} colour combinations, and at around 135° degrees, representing \textit{triadic} relationships. \textit{Complementary} colours (180° degrees apart) are traditionally considered harmonious, but this preference appears to manifest only within US and French responses.

For musical harmonies, we generally replicate the patterns previously observed among the US and Korean participants using the same experimental design~\cite{marjieh_timbral_2024}. Our analysis also reveals strong preferences for \textit{unison} (same tone), \textit{perfect fifth} (7 semitones), and \textit{octaves} (12 semitones). We also find notable dips in \textit{tritone} (6 semitones), which is known to be displeasing. However, some countries deviate from these established patterns. In Mexico, for instance, the tritone is not particularly disliked. Additionally, countries such as India and Nigeria show more uniform preferences (i.e., flatter lines) across intervals.

For melodies, the defined space was identical to harmonies. However, we find strikingly different patterns with substantial cross-cultural variation. Notably, for some countries (e.g., Mexico), preference peaks do not align closely with the standardised pitch intervals (e.g., keys on a piano). Yet, most countries demonstrate tendencies for favouring octave and disfavouring tritone melodies, which is in line with past observation on US and Indian online participants~\cite{anglada-tort_large-scale_2023}.

\subsection{Agreement and Disagreement Between Cultures}
Finally, we empirically quantified the extent of shared agreement on aesthetic preferences between cultures. For all pairwise combinations of countries in each modality, we computed Spearman correlations for the entire preference spaces (Figure~\ref{fig:fig4}A), followed by computing correlations between GAM fittings that describe mathematical relational preferences (Figure~\ref{fig:fig4}B).

High or low between-country correlations can be influenced by the reliability, or noisiness, of each modality. We thus added to the bottom of Figure~\ref{fig:fig4} that reports within-modality reliability by computing split-half correlations within each country with 100 bootstrap simulations, and then aggregating to calculate the mean.

Our analysis shows that preferences for shapes (space rho = 0.84, 95\% CI = [0.83, 0.86]; relational rho = 0.89 [0.88, 0.91]) and curvatures (space rho = 0.77 [0.73, 0.81]; relational rho = 0.80 [0.75, 0.85]) are generally highly agreed across cultures.

For colour, between-country correlations are high when comparing spaces (rho = 0.84 [0.82, 0.86]), which stems from most countries favouring colour combinations with bluish hues (as previously seen in Figure~\ref{fig:fig2}). By contrast, there exists little agreement for relational differences in hue degrees (rho = 0.46 [0.35, 0.57]). This suggests that colour preferences are more absolute and categorical.

An opposite pattern emerges for musical harmony. While there is little agreement between spaces (rho = 0.28 [0.21, 0.34]), we see strong agreement in preferred musical intervals (rho = 0.78 [0.75, 0.80]). This demonstrates that harmony preferences follow certain ratio rules~\cite{helmholtz_sensations_1954}, whereas their spatial preferences can vary (e.g., as seen in Figure~\ref{fig:fig2}B, Germans preferred harmonies lower in the frequency spectrum). By contrast, melody shows consistently the lowest agreement between countries (space rho = 0.25 [0.19, 0.30], relational rho = 0.38 [0.31, 0.46]).

It is important to note, however, modalities with little agreement across cultures generally also have low reliability (i.e., low within-country agreement). As such, modalities with lower agreement may underlie either (i) noisier cognitive processing for evaluating preferences in those modalities, or (ii) there being higher variability in individual preferences.


\section{Discussion}
Our large-scale cross-cultural study captures diverse cultural nuances in aesthetic preferences worldwide. We found cultural variation in almost all modalities. Ratio relationships within each modality were almost always important, except for colours, where we observed categorical behaviour that are consistent with previous findings (e.g.,~\citeNP{vangeert_jacoby_2024}). Furthermore, the amount of variability between cultures was substantially different across the modalities: for instance, shape and curvature showed more universal preference, while melodic preferences were highly varied between countries. These findings together highlight how certain aspects of aesthetic appreciation might be less influenced by cultural learning, rather, driven by a psychological foundation that is more rooted in human biology.

Conducting the study online enabled broad scalability, but it comes with the cost of some experimental control. Traditional laboratory studies of colour perception utilise precisely calibrated monitors, and music experiments take place in acoustically isolated environments. Online participants are also influenced by global and mainstream media. Future work should replicate the paradigm in laboratory settings with participants from diverse cultures including small-scale societies~\cite{jacoby_universal_2019, Jacoby2024-tw}. Moreover, the specified parameter spaces we used could also have missed other important areas of cultural nuances. Hence, they should be extended to explore different parameter ranges, dimensions (e.g., varying lightness instead of colour hue), and higher dimensions using efficient sampling methods~\cite{harrison2020, Van_Rijn2022-ol}.

Our use of simplified, two-dimensional stimuli allowed for systematic exploration of aesthetic preferences across modalities and cultures. However, it may have been insufficient to fully capture the complexity of aesthetic experiences in naturalistic settings (e.g., paintings and songs), and future works can incorporate complex stimuli to enhance ecological validity.

To conclude, our comprehensive study to understand global aesthetic preference reveals rich and complex cultural variations. Accordingly, it opens research avenues on the mechanisms underlying this variability, from demographic compositions~\cite{Lee2024-yo}, and emotional associations~\cite{Lee2021-zb}, to the influence of globalisation~\cite{pieterse_2025}. Such insights can have broad implications in cognitive science, social science, psychology, and empirical aesthetics.

\bibliographystyle{apacite}

\setlength{\bibleftmargin}{.125in}
\setlength{\bibindent}{-\bibleftmargin}

\bibliography{CogSci_Template}

\end{document}